\documentclass[fleqn,usenatbib]{mnras}

\usepackage{newtxtext,newtxmath}
\usepackage[T1]{fontenc}

\DeclareRobustCommand{\VAN}[3]{#2}
\let\VANthebibliography\thebibliography
\def\thebibliography{\DeclareRobustCommand{\VAN}[3]{##3}\VANthebibliography}

\usepackage{graphicx}	
\usepackage{amsmath}	
\usepackage{multirow}

\newcommand{\meth}   {CH$_3$OH}
\newcommand{\kms}    {km\,s$^{-1}$}

\newcommand{\tgas}    {$T_{\rm gas}$}

\title[CH$_3$OH, HCN and CO in C/2017~K2]{Single-dish observations and non-LTE analysis of CH$_3$OH, HCN, and CO line emission in the Oort cloud comet C/2017 K2 (PANSTARRS)}

\author[M. S. Kirsanova et al.]{
M. S. Kirsanova,$^{1}$\thanks{E-mail: kirsanova@inasan.ru (MSK)}
Ya. N. Pavlyuchenkov$^{1}$
A. O. H. Olofsson,$^{2}$
M. S. Lerner$^{2}$\\
$^{1}$Institute of Astronomy of the Russian Academy of Sciences, Pyatnitskaya str. 48, 119017, Moscow, Russia\\
$^{2}$Department of Physics and Astronomy, Chalmers University of Technology, Onsala Space Observatory, SE-439 92 Onsala, Sweden
}

\date{Accepted XXX. Received YYY; in original form ZZZ}

\pubyear{2015}

\begin{document}
\label{firstpage}
\pagerange{\pageref{firstpage}--\pageref{lastpage}}
\maketitle

\begin{abstract}
We present pre-perihelion observations of methanol, carbon monoxide, and hydrogen cyanide in the Oort cloud comet C/2017 K2 (PANSTARRS), performed with the APEX 12-m and Onsala 20-m telescopes from April to July 2022. As the comet’s heliocentric distance decreased from 3.4 to 2.7~AU, CH$_3$OH line intensities increased substantially (by factors of 1.1–4.0), with the most pronounced enhancement in lines with upper-level energies $E_{\rm u} > 40$~K. In contrast, the brightness of the CO and HCN lines remained constant. We estimate the best-fit gas kinetic temperatures $T_{\rm gas} >100 $~K and water production rate of $Q_{\rm H_2O}=(3-10)\times10^{28}$~s$^{-1}$. The derived methanol-to-water abundance ratio $\approx 0.01-0.04$, depending on the observed period. Our results demonstrate that non-LTE effects are dominant in the coma and must be accounted for to accurately derive molecular production rates. We also report weak non-thermal excitation, including potential maser activity in the CH$_3$OH $8_0-7_1$ line.
\end{abstract}




\section{Introduction}

Advances in technology are leading to the discovery of a growing number of objects beyond Neptune. They are also increasing the list of active comets found outside Jupiter's orbit \citep[see e.~g.][]{1995AJ....109.1867J}; see also the study by \citet{2023A&A...676A.104F} and the CODE catalog \citep{2025arXiv250820780D}. Consequently, comet researchers have concentrated their efforts on studying these trans-Neptunian bodies as a new class of comet-like objects. Studying them provides insight into the origin and formation of the Solar System. 

While the activity of comets within 3~AU can be explained by sublimation of water ice heated by solar radiation \citep[see  e.~g.][]{2011ARAA..49..471M}, the activity of distant comets must be related to sublimation of more volatile species like CO, CO$_2$, or N$_2$ \citep[see e.~g.][]{2017PASP..129c1001W}. Cometary water ice can only form in an environment with a temperature of $T \leq 25-30$~K. Therefore, comet nuclei must have formed and been covered with water ice early in the Solar System's history—either in the protoplanetary disk or in the parent molecular cloud. This scenario suggests that comet nuclei are as old as the Solar System itself. 

A detailed study of the Oort cloud comets is also necessary to understand the formation of giant planet atmospheres. Current observational data, including data from the Galileo space telescope, indicate a high abundance of metals relative to hydrogen in these atmospheres \citep{2003NewAR..47....1Y}. These measured values are significantly higher than Solar abundances. To explain this, models of giant planet formation require an additional source of material probably accreted from external sources \citep{2022Icar..37814937H}. This material could have been delivered by Oort Cloud comets, that formed at temperatures below 30 K beyond 30~AU from the young Sun~\citep{2002ESASP.500..753M}.

Distant comets, as members of the Oort and Kuiper belts, represent some of the most primordial material in the Solar System. Because they orbit at large heliocentric distances, their material remains in a largely unprocessed state. The icy surfaces of comet nuclei contain a variety of molecules, ranging from simple diatomic compounds to complex organic precursors like methanol \citep[e.~g.][]{2002AA...390..363I}. Measuring the molecular composition of distant comets helps determine the extent to which their ices have been processed by solar UV radiation. Furthermore, comparing the composition of cometary and interstellar ices reveals how much interstellar matter survived and was incorporated into the protosolar nebula \citep[e.g.][]{2019MNRAS.490...50D, 2024ApJ...970L...5L, 2025AJ....170..312W}.

Since the main channel of methanol formation is the hydrogenation of CO molecules accreted from the gas phase on the grains \citep{2003ApJ...588L.121W, 2005IAUS..231..237C}, the abundance ratio of the primary volatile molecules CH$_3$OH/CO can serve as a marker of a comet's history and its exposure to solar UV photons. For instance, \citet{2017AA...603A.124C} demonstrate that methanol and CO can be abundant even in harsh environments like the Orion Nebula, where the UV field is as strong as $10^5$ times the average interstellar value. Similarly, \citet{2021MNRAS.507.3810K} find that interstellar abundances of CH$_3$OH, formed during the dark stage of a molecular cloud, can be preserved for a long time, even near young stars, thereby revealing the history of star formation. Therefore, monitoring CO (as a precursor to CH$_3$OH) and CH$_3$OH line emissions in comets as they approach perihelion allows us to estimate how much pristine interstellar material survives and to quantitatively measure how solar irradiation alters the CH$_3$OH/CO ratio.

The primary goal of this study is to explore the CH$_3$OH, CO, and HCN line emissions from the Oort cloud comet C/2017 K2 (PANSTARRS) to determine the abundances of these molecules in its ices. This comet is a unique target, having shown significant activity as far out as 23 au \citep{2019AJ....157...65J}. With a perihelion of 1.79~au, our study tracks how its activity changed under different temperature conditions—from beyond the water ice sublimation zone to closer distances where water ice sublimation dominates. Our observations from April to July 2022 captured the comet at 2.7–3.4~AU from the Sun, where the sublimation of CO$_2$ and H$_2$O ices competes. We aim to analyze this unique object before it reaches perihelion and crosses the water-ice sublimation line. This article concentrates on methanol, given the numerous spectral lines we observed for this molecule. Since the time of our observations, several studies dedicated to C/2017~K2 have appeared, including some using interferometers (see the discussion in Sec.~\ref{sec:disc}). Consequently, we focus our analysis on a comparison of LTE and non-LTE methods as applied to the observed methanol lines in this article.

\section{Observations}\label{sec:obs_mm}

We observed C/2017~K2 with the APEX 12-m telescope in April and July of 2022 focusing on the methanol $5_K-4_K$ series at 240~GHz and the CO(2-1) line at 231~GHz within Swedish operation time. The receiver used was the nFLASH230 with a spectral resolution of 61~kHz or $0.08$~km~s$^{-1}$ at the frequency of the methanol lines. With the receiver tuned to 230~GHz in the LSB, we also obtained the methanol series in the USB and observed these lines simultaneously. The observations were performed using the Wobble observational mode with the throw of 300\arcsec. The integration time was 30~sec. The pointing was checked within a hour between them. We found the pointing RMS $<2$\arcsec{} all over the sky.

Observations of C/2017~K2 with Onsala Space Observatory (OSO) 20-m telescope have been performed using the 3/4-mm receiver which is a dual-polarization, dual-sideband receiver \citep{Belitsky_2015}. The backend consists of four 32768-channel FFT spectrometers with a maximum bandwidth of 2.5~GHz each and a spectral resolution of 76.3~kHz/channel. Most comet measurements have been performed in the ``4~GHz mode'' where two pairs of spectrometers are used together with a frequency shift to cover 4~GHz bandwidth in total (with a 1~GHz overlap). Some observations of CH$_3$OH $2_K-1_K$ series have used a dual sideband mode where 2.5~GHz is recorded from the lower sideband simultaneously with 2.5~GHz from the upper sideband. 

The observations have been performed in frequency switching mode (FSW), where the observed frequency jumps between two values with a 5~Hz period. The frequency throw used has been 7~MHz for HCN(1--0) and 25~MHz for CH$_3$OH $2_K-1_K$ series, with the throw selected to avoid overlap between spectral lines when folding the spectra. The receiver is known to provide stable FSW observations with at least 30~MHz throw, but baselines are getting worse the larger the throw is, but that is typically not a problem for this study, since the cometary spectral lines are quite narrow. About 6.7\% of the observing time is lost due to blanking when FSW is used. Pointing and focusing are done by observing the SiO line at 86.243~GHz from some strong point source. The pointing RMS was $2-3$\arcsec{} all over the sky.

The beam size of the telescopes varies with frequency: from  42.9\arcsec{} for the HCN(1--0) line to 39.3\arcsec{} and 26\arcsec{} for the methanol $2_K-1_K$ and $5_K-4_K$ series simultaneously. The noise level of our data is $\approx 5-10$~mK in the $T_{\rm mb}$ scale, depending on the date for the selected spectral resolution. 

The position of the comet is taken from ephemeris files obtained from JPL Horizons \citep[][]{1996DPS....28.2504G}. Positions and relative velocities are tabulated at 1-hour intervals, and the telescope control system is using a second order polynomial to continuously interpolate the position. The frequency tunings were performed in the velocity frame of the comet using its velocity relative to the observer from the JPL Horizons and the velocity of the observer relative to the local standard of rest.

Data were reduced and analyzed with the GILDAS\footnote{http://www.iram.fr/IRAMFR/GILDAS} software.

\section{Observational results}\label{sec:obsresults}

\begin{figure*}
    \centering
    \includegraphics[width=2\columnwidth]{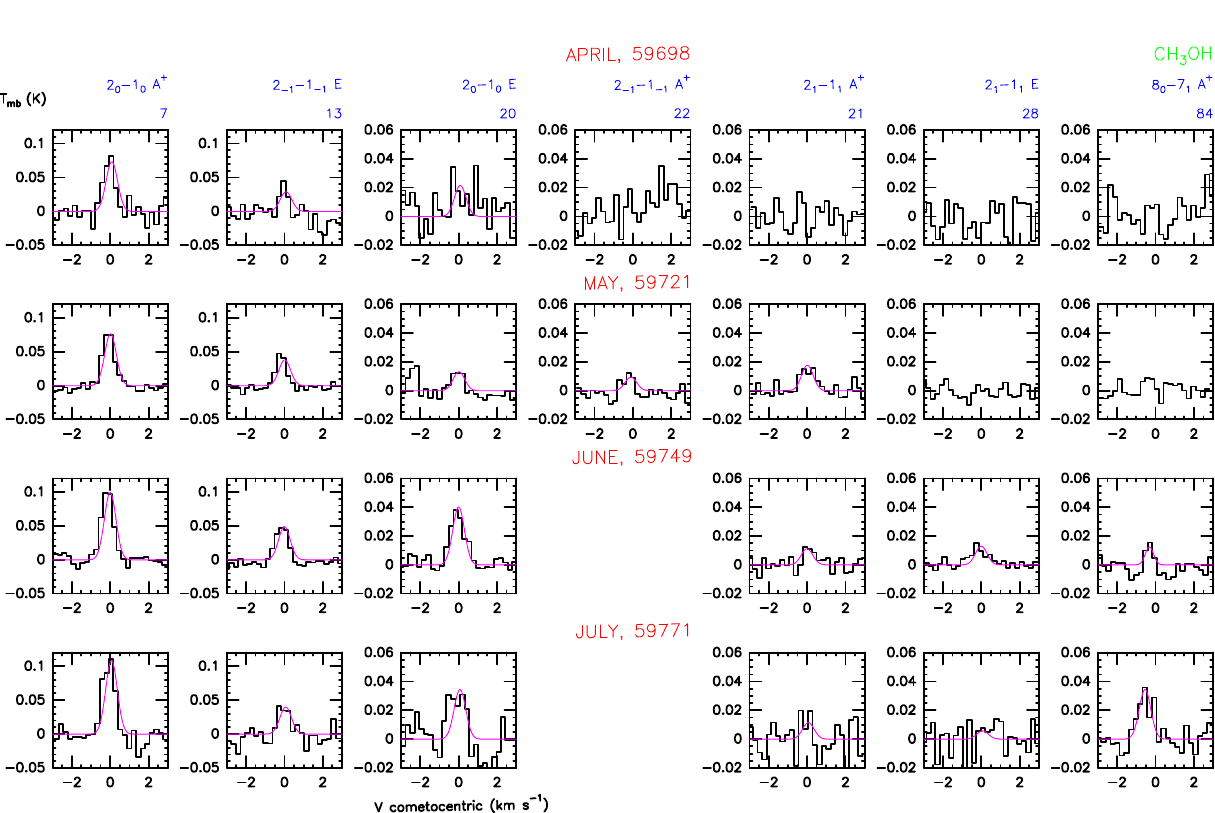}\\
    \caption{Spectra of methanol at 96~GHz obtained with the 20-m OSO telescope. The Gaussian fit from the Table~\ref{sec:obsresults} is shown by magenta. Modified Julian Date is shown by red along with the month the spectra were obtained. Quantum numbers of the observed spectral lines along with the $E_{\rm u}$ values are shown by blue. }
    \label{fig:apr_96_241}
\end{figure*}

\begin{figure*}
    \centering
    \includegraphics[width=1.8\columnwidth]{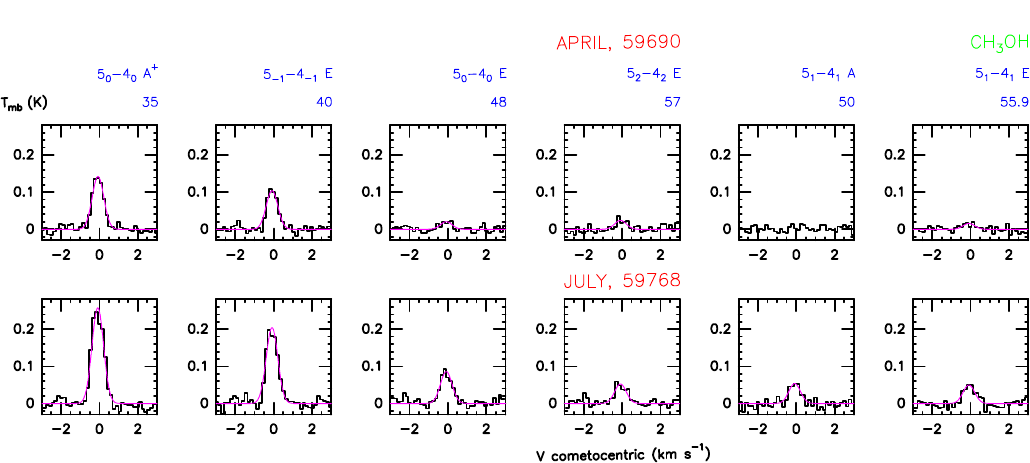}
    \caption{Spectra of methanol at 241~GHz obtained with the APEX telescope at two observational periods. The labelling is the same as in Fig.~\ref{fig:apr_96_241}. }
    \label{fig:july_96_241}
\end{figure*}

Our observations from April to July covered the range of the Modified Julian Date from 59690 to 59771 corresponding to the pre-perihelion heliocentric distances $r_{\rm h}$ from 3.4 to 2.7~AU. It was a period of the closest approach of the comet to the Earth, with the geocentric distances $\Delta$ from 3.1 to 1.8~AU, respectively. Detected spectral lines of CH$_3$OH, HCN, and CO molecules on particular observing days and the corresponding orbital elements are given in Table~\ref{tab:detected}. For the subsequent analysis, we fitted all the methanol lines by the Gauss profiles with the same width for each particular day. Fig.~\ref{fig:apr_96_241}-\ref{fig:july_96_241} shows the detected lines of methanol emission over the observed period.

We will refer to the data obtained in April and July as the 'spring' and 'summer' periods hereafter and will use these data for most of the analysis, since the two methanol line series were obtained in these months. Approaching the Sun and subsequently the Earth from April to July, the comet revealed more bright emission of the methanol lines. Namely, lines of the $2_K-1_K$ series with $E_{\rm u} = 6.9$~and~12.5~K became brighter by 1.1-1.2 times. The line with $E_{\rm u}=20.1$~K became more than twice as bright in June and July~2022 compared with the spring. The rest of the lines in the series were not visible at all, at the noise level of 12~mK, in April or May. The lines of the $5_K-4_K$ series enhanced their brightness even more significantly. The lines with $E_{\rm u} \leq 40$~K became brighter by 1.8-1.9 times. The rest of the lines have $E_{\rm u} > 40$~K, their brightness enhanced by 2-4 times.

\begin{figure*}
    \centering
    \includegraphics[width=1.5\columnwidth]{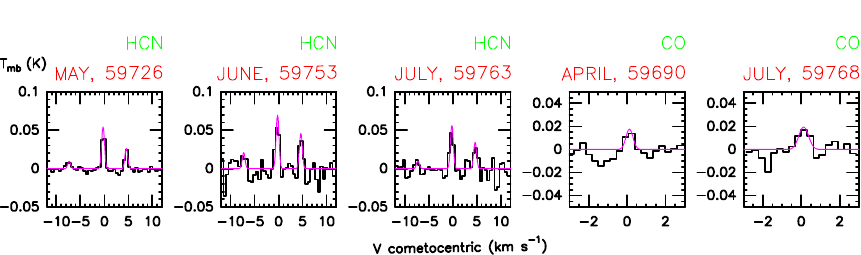}
    \caption{HCN(1--0) and CO(2--1) line emission at three observational periods. The labelling is the same as in Fig.~\ref{fig:apr_96_241}. }
    \label{fig:HCN}
\end{figure*}

The observed lines of CO and HCN molecules revealed different behaviour — their brightness did not rise from April to July, as shown in Fig.~\ref{fig:HCN}. To illustrate how the intensities change, we show their values in an additional figure (Fig.~\ref{fig:julian_date}) and compare with the methanol lines. It can be seen that the intensities of the HCN line emission obtained in May, June, and July agree with each other within the uncertainties. While the line intensities appear higher in summer compared to spring, the lower signal-to-noise ratio of the summer spectra does not allow for a confident conclusion. The intensities of the CO(2--1) line remain certainly the same.

The widths of the methanol and HCN lines were the same during the entire observed period $\Delta v = 0.715\pm0.01$~\kms, where the latter value is the formal uncertainty of the Gauss fit. The detected CO(2--1) line was $\approx 1.6$~times narrower in April and became as broad as the rest of the lines in July. As most of the observed lines have the same width, we obtain the velocity of the coma expansion of $=0.3$~\kms{} which was unchanged regardless of the position of the comet relative to the Sun.

\begin{table*}
\centering
\caption{Detected molecular lines towards the comet in April and July 2022. The values $r_{\rm h}$, $\Delta$ and $d_{\rm min}$ are heliocentric distance, geocentric distance of the comet and the minimum distance of the sampled coma from the nucleus, respectively. Modifies Julian Date (MJD) is given for the each particular day. $^*$Velocities and widths of the different $K$ lines of CH$_3$OH are assumed to be the same in the fitting. $^{**}$The upper limits of the line brightnesses correspond to $2\sigma$. Spectral resolution was degraded to 0.3~\kms{} to detect the CO(2--1) line.}
\begin{tabular}{lccccccc}
\hline
Sym                         & Transition                                &
Frequency                   & $E_{\rm u}$& $\int T_{\rm mb} dV$ &$V$ &$\Delta v$   & T$_{\rm mb}$ \\
                            &                                           &
[MHz]                      & [K]            & [mK\,\kms]            & [\kms]       &[\kms]       & [mK]          \\
\hline
\multicolumn{8}{c}{21 April 2022, MJD=59690, $r_{\rm h}=3.4$~AU, $\Delta = 3.1$~AU, $d_{\rm min} = 2.8\times10^4$~km} \\
\multicolumn{8}{c}{\bf \meth}\\
E    & $5_0-4_0 $      & 241700.159 & 47.9 & $16\pm 4$  &  -"-            &  -"-            & $21\pm 13$ \\
E    & $5_{-1}-4_{-1}$ & 241767.234 & 40.4 & $80\pm 4$  &  -"-            &  -"-            & $104\pm 13$ \\
A$^+$& $5_{0}-4_{0}$   & 241791.352 & 34.8 & $109\pm 4$ & $-0.07\pm 0.01$ & $0.715\pm 0.01$ & $144\pm 13$ \\
E    & $5_{1}-4_{1}$   & 241879.025 & 55.9 & $12\pm 4$  &  -"-            &  -"-            & $16\pm 13$ \\
E    & $5_{2}-4_{2}$   & 241904.643 & 57.0 & $18\pm 4$  &  -"-            &  -"-            & $24\pm 13$ \\
A       & $5_{1}-4_{1}$   & 243915.788  & 49.7        &              &            &             & $<26$   \\
\multicolumn{8}{c}{\bf CO} \\
  -- & $2-1$           & 230538.000 & 16.6 & $10\pm 3$ & $0.11\pm 0.06$ & $0.450\pm 0.132$ & $22\pm 7$ \\
\multicolumn{8}{c}{ }\\
\multicolumn{8}{c}{29 April 2022, MJD=59698, $r_{\rm h}=3.3$~AU, $\Delta = 2.9$~AU, $d_{\rm min} = 3.9\times10^4$~km}\\
\multicolumn{8}{c}{\bf \meth}\\
A    &   $2_0-1_0 $      & 96741.371 & 6.9  & $68\pm6$ & -"-             & -"-             & $90\pm8$\\
E    &   $2_0-1_0 $      & 96744.545 & 20.1 & $14\pm6$  & -"-             & -"-             & $18\pm8$\\
E    &   $2_{-1}-1_{-1}$ & 96739.358 & 12.5 & $37\pm6$ & $0.087\pm0.030$ & $0.715\pm0.01$ & $48\pm8$\\
E    &   $2_1-1_1 $      & 96755.501 & 28.0 & $-$  & -"-             & -"-             & $<23^{**}$\\
A    &   $2_1-1_1 $      & 95914.310 & 21.4 & $-$ & -"-             & -"-             & $<23$\\
A    &   $8_0-7_1 $      & 95169.391 & 83.5 & $-$  & -"-             & -"-             & $<23$\\
A    &   $2_{-1}-1_{-1}$ & 97582.798 & 21.6 & $-$  & -"-             & -"-             & $<23$\\
\multicolumn{8}{c}{ }\\
\multicolumn{8}{c}{22 May 2022, MJD=59721, $r_{\rm h}=3.1$~AU, $\Delta = 2.4$~AU, $d_{\rm min} = 3.2\times10^4$~km}\\
\multicolumn{8}{c}{\bf \meth}\\
A    &   $2_0-1_0 $      & 96741.371 & 6.9  & $67\pm2$ & -"-             & -"-             & $87\pm5$\\
E    &   $2_0-1_0 $      & 96744.545 & 20.1 & $9\pm2$  & -"-             & -"-             & $12\pm5$\\
E    &   $2_{-1}-1_{-1}$ & 96739.358 & 12.5 & $38\pm2$ & $0.024\pm0.014$ & $0.715\pm 0.01$ & $50\pm5$\\
E    &   $2_1-1_1 $      & 96755.501 & 28.0 & $1\pm2$  & -"-             & -"-             & $<10$\\
A    &   $2_1-1_1 $      & 95914.310 & 21.4 & $12\pm2$ & -"-             & -"-             & $16\pm5$\\
A    &   $8_0-7_1 $      & 95169.391 & 83.5 & $4\pm2$  & -"-             & -"-             & $<10$\\
A    &   $2_{-1}-1_{-1}$ & 97582.798 & 21.6 & $7\pm4$  & -"-             & -"-             & $6\pm2$\\
\multicolumn{8}{c}{ }\\
\multicolumn{8}{c}{27 May 2022, MJD=59726, $r_{\rm h}=3.1$~AU, $\Delta = 2.3$~AU, $d_{\rm min} = 3.3\times10^4$~km}\\
\multicolumn{8}{c}{\bf HCN}\\
--   &   $1_{1}-0_{1}$   & 88630.416 & -"- & $22\pm 2$  & -"-             & -"-             & $29\pm 4$\\
--   &   $1_{2}-0_{1}$   & 88631.847 & 4.3 & $44\pm 2$  & $-0.273\pm0.019$  & $0.715\pm 0.01$ & $59\pm 4$\\
--   &   $1_{0}-0_{1}$   & 88633.936 & -"- & $8\pm 2$  & -"-             & -"-             & $12\pm 4$\\
\multicolumn{8}{c}{ }\\
\multicolumn{8}{c}{ 19 Jun 2022, MJD=59749, $r_{\rm h}=2.9$~AU, $\Delta = 1.9$~AU, $d_{\rm min} = 2.6\times10^4$~km }\\
\multicolumn{8}{c}{\bf \meth}\\
A    &   $2_0-1_0 $      & 96741.371 & 6.9  & $79\pm3$ & -"- & -"- & $105\pm5$\\
E    &   $2_0-1_0 $      & 96744.545 & 20.1 & $33\pm3$ & -"- & -"- & $43\pm5$\\
E    &   $2_{-1}-1_{-1}$ & 96739.358 & 12.5 & $42\pm3$ & $0.002\pm 0.013$ & $0.715\pm 0.01$ & $56\pm5$\\
E    &   $2_1-1_1 $      & 96755.501 & 28.01& $18\pm3$ & -"- & -"- & $23\pm5$\\
A    &   $2_1-1_1 $      & 95914.310 & 21.4 & $7\pm3$ & -"- & -"- & $8\pm5$\\
A    &   $8_0-7_1 $      & 95169.391 & 83.5 & $16\pm2$ & -"- & -"- & $22\pm5$\\
\multicolumn{8}{c}{ }\\
\multicolumn{8}{c}{23 Jun 2022, MJD=59753, $r_{\rm h}=2.8$~AU, $\Delta = 1.9$~AU, $d_{\rm min} = 2.8\times10^4$~km}\\
\multicolumn{8}{c}{\bf HCN}\\
--   &   $1_{1}-0_{1}$   & 88630.416 & -"- & $40\pm 10$  & -"-             & -"-             & $53\pm 18$\\
--   &   $1_{2}-0_{1}$   & 88631.847 & 4.3 & $55\pm 10$  & $-0.271\pm0.068$  & $0.715\pm 0.01$ & $72\pm 18$\\
--   &   $1_{0}-0_{1}$   & 88633.936 & -"- & $-$         & -"-             & -"-             & $<36$\\
\multicolumn{8}{c}{ }\\
\multicolumn{8}{c}{03 Jul 2022, MJD=59769, $r_{\rm h}=2.8$~AU, $\Delta = 1.8$~AU, $d_{\rm min} = 2.7\times10^4$~km}\\
\multicolumn{8}{c}{\bf HCN}\\
--   &   $1_{1}-0_{1}$   & 88630.416 & -"- & $28\pm 8$  & -"-             & -"-             & $36\pm 14$\\
--   &   $1_{2}-0_{1}$   & 88631.847 & 4.3 & $46\pm 8$  & $-0.226\pm0.068$  & $0.715\pm 0.01$ & $61\pm 14$\\
--   &   $1_{0}-0_{1}$   & 88633.936 & -"- & $-$        & -"-             & -"-             & $<28$\\
\hline
\end{tabular}
\label{tab:detected}
\end{table*}

\begin{table*}
\centering
\contcaption{Table~1}
\begin{tabular}{lccccccc}
\hline
Sym     & Transition      & Frequency   & $E_{\rm u}$ & $\int T_{\rm mb} dV$ & $V$   & $\Delta V$      & $T_{\rm mb}$ \\
        &                 & [MHz]       & [K]         & [mK\,\kms]           & [\kms]          & [\kms]          & [mK]         \\
\hline
\multicolumn{8}{c}{8 July 2022, MJD=59768,  $r_{\rm h}=2.7$~AU, $\Delta = 1.8$~AU, $d_{\rm min} = 1.7\times10^4$~km}\\
\multicolumn{8}{c}{\bf \meth}\\
E       & $5_0-4_0 $      & 241700.159  & 47.9        & $66\pm 4$            & -"-             &  -"-            & $86\pm 11$  \\
E       & $5_{-1}-4_{-1}$ & 241767.234  & 40.4        & $157\pm 4$           & -"-             &  -"-            & $206\pm 11$ \\
A$^+$   & $5_{0}-4_{0}$   & 241791.352  & 34.8        & $199\pm 4$           & $-0.08\pm 0.01$ & $0.715\pm 0.01$ & $261\pm 11$ \\
E       & $5_{1}-4_{1}$   & 241879.025  & 55.9        & $37\pm 4$            & -"-             &  -"-            & $49\pm 11$  \\
E       & $5_{2}-4_{2}$   & 241904.643  & 57.0        & $40\pm 4$            & -"-             &  -"-            & $52\pm 11$  \\
A       & $5_{1}-4_{1}$   & 243915.788  & 49.7        & $41\pm4$             & -"-             &  -"-            & $54\pm$11   \\
\multicolumn{8}{c}{ }\\
\multicolumn{8}{c}{\bf CO}\\
  --    & $2-1$           & 230538.000  & 16.6        & $13\pm 4$            & $0.13\pm 0.11$  & $0.656\pm 0.183$& $18\pm 6$   \\
\multicolumn{8}{c}{ }\\
\multicolumn{8}{c}{11 July 2022, MJD=59771, $r_{\rm h}=2.7$~AU, $\Delta = 1.8$~AU, $d_{\rm min} = 2.4\times10^4$~km}\\
\multicolumn{8}{c}{\bf \meth}\\
A       &   $2_0-1_0 $      & 96741.371 & 6.9         & $88\pm5$             & -"-             &  -"-            & $116\pm10$\\
E       &   $2_0-1_0 $      & 96744.545 & 20.1        & $32\pm5$             & -"-             &  -"-            & $42\pm10$ \\
E       &   $2_{-1}-1_{-1}$ & 96739.358 & 12.5        & $34\pm5$             & $0.070\pm 0.024$& $0.715\pm 0.01$ & $44\pm10$ \\
E       &   $2_1-1_1 $      & 96755.501 & 28.01       & $14\pm5$             & -"-             &  -"-            & $18\pm10$ \\
A       &   $2_1-1_1 $      & 95914.310 & 21.4        & $8\pm6$              & -"-             &  -"-            & $11\pm10$ \\
A       &   $8_0-7_1 $      & 95169.391 & 83.5        & $23\pm4$             & $-0.50\pm 0.02$&  -"-            & $31\pm10$ \\
\hline
\end{tabular}
\end{table*}

\section{Preliminary LTE analysis}

We start with the preliminary LTE analysis of methanol emission using the rotation diagram method \citep[see][for all details regarding to the methanol lines]{Kalenskii2016}. Using this approach, we used the brightness temperatures of the observed lines to estimate the column densities of the emitting molecules. Since we observed at different frequencies, we need to account for the beam filling factors at each frequency. Assuming we are dealing with an unresolved source with a uniform brightness temperature, the ratio of the antenna temperature (on the main beam scale, $T_{\rm mb}$ in our case) to the brightness temperature is defined as the ratio of the source solid angle to the beam solid angle. As the telescope beams at 96 and 241~GHz are different and the filling factor is generally unknown, we tried several values, fitting the column densities at the upper level divided by the statistical weight of the level ($N_{\rm u}/g_{\rm u}$) on the diagrams. The values of $g_{\rm u} = 11, 5$ and 17 for the methanol lines of the $5_K-4_K$, $2_K-1_K$ series and the $8_0-7_1$ line, respectively. The best agreement to the straight line we obtained assuming the filling factor $f_{241}=1$ at 241~GHz and $f_{96}=0.41$ at 96~GHz, where the latter is the ratio of squared beam angular sizes $\Theta_{241}^2/\Theta_{96}^2$. Assuming $f_{96}=f_{241}=1$ we obtain worse agreement to the straight line, but the values for the rotational temperatures $T_{\rm rot}$ are similar. Namely, we obtain $T_{\rm rot}=11.3\pm1.0$ and $T_{\rm rot}=14.0\pm0.9$~K for spring and summer, respectively, with $f_{96}=0.41$. For $f_{96}=1$, we obtain $T_{\rm rot}=13.5\pm1.8$ and $T_{\rm rot}=19.9\pm1.9$~K for spring and summer, respectively. 

From this preliminary analysis we find the excitation temperature for the methanol emission remained stable or changed insignificantly, even as the comet traversed the water snow line and its heliocentric distance decreased by 25\% between spring and summer. Methanol column density ($N_{\rm CH_3OH}$) increased by 1.6 times from $N_{\rm CH_3OH}=(1.1\pm0.1)\times 10^{13}$~cm$^{-2}$ to $(1.8\pm0.2)\times 10^{13}$~cm$^{-2}$ between spring and summer for the solution with $f_{96}=0.41$. We also note the $N_{\rm u}/g_{\rm u}$ values for the E-type methanol better agree with the linear fit line than the values for the A-type methanol in the summer diagrams and discuss this non-LTE effect in the next section.

\section{Advanced analysis}

After the preliminary LTE analysis, we performed non-LTE calculations using the physical model of the comet from \citet{Bensch_2004} and radiation transfer code URAN(IA) by~\citet{2004ARep...48..315P, 2008ApJ...689..335P}. We summarise the model assumptions and describe the process of the modelling below. 

\subsection{Model}

We approximate the radial distribution of molecules in the coma by spherically symmetric Haser model \citep{1957BSRSL..43..740H}:

\begin{equation}
n(r)=\frac{Q}{4 \pi r^2 v_{\rm exp}} \exp{ \left( -\frac{r \beta}{v_{\rm exp}} \right) },
\end{equation}

where $Q$ is the molecular production rate, $r$ is cometocentric distance, $\beta$ is the photodissociation rate at particular heliocentric distance $r_{\rm h}$ and $v_{\rm exp}$ is the isotopic expansion of the coma. Since the cometary coma consists mainly of water, this molecule will be the main collisional partner for methanol at least in the inner coma. Moreover, the cometocentric radial distribution of water is needed to set up the distribution of electron density, which is the main collision partner for the external part of the coma (see below). Our observations are single-dish and we cannot study possible differences in the spatial distributions of methanol and water. Therefore, we model water radial distribution and use constant ratio of CH$_3$OH/H$_2$O abundances in calculations.

The photodissociation rates for molecules were estimated for the particular values of $r_{\rm h}$ in the same way as it was done for water by \citet{1994Icar..107..164B}: $\beta = \beta_0/r_{\rm h}^2$, where $r_{\rm h}$ is in AU and $\beta_0 = 1.042\times10^{-5}$~s$^{-1}$ is the water photodissociation rate at 1~AU. Since the photodissociation rates for water and methanol differ only by 9\% in the Solar radiation field \citep[][]{2017A&A...602A.105H}, and they both are parent molecules, we assume the same rates and radial distribution for water and methanol but vary only the abundance ratio of these molecules as $Q_{\rm CH_3OH} = x Q_{\rm H_2O}$, where $x<1$. Below we designate $x$ as ${\rm CH_3OH/H_2O}$. 

We assumed constant gas temperature (\tgas) for the neutral molecules throughout the coma and check if the constant \tgas{} below $\sim 100$~K in the region where neutral gas collisions are important is the best way to fit the observed line intensities as was found by \citet{Bensch_2004}.

Following the cited work and the measurements in 1P/Halley we model the radial distribution of the electron density $n_{\rm e}(r)$ by the following equation:
\begin{equation}
\begin{split}
n_{\rm e}(r) = x_{\rm n_e} \left( \frac{Q_{\rm H_2O} k_{\rm ion}}{r^2_{\rm h} v_{\rm exp} k_{\rm rec} } \right)^{0.5}  \left(  \frac{T_{\rm e}}{300{\rm K}}\right)^{0.15}  \\\left(   \frac{R_{\rm rec}}{r^2} \right)   \left[  1-\exp\left(-\frac{r}{R_{\rm rec}}  \right)   \right],
\end{split}
\end{equation}

where $R_{\rm rec}=3200 x_{\rm r_e} Q_{29}^{1/2}$~km is the recombination surface. The recombination rate of electrons with ions is insignificant in the region where $r>R_{\rm rec}$, $Q_{29}=Q_{\rm H_2O}/10^{29}$, $x_{\rm r_e}=1$. The photoionization rate $k_{\rm ion}=4.1\times10^{-7}$~s$^{-1}$ and the recombination rate of ions with electrons $k_{\rm rec}=(300 {K}/T_{\rm e})^{1/2} 0.7\times 10^{-6}$~s$^{-1}$~cm$^3$. The value $x_{\rm n_e}=1$ is the scaling coefficient and we use this calibration from \citet{Bensch_2004}.

We use the three different approaches for the electron temperature $T_{\rm e}$ inside, around, and outside the contact surface ($R_{\rm cs}$). Namely, $T_{\rm e}=T_{\rm gas}$ if $r<R_{\rm cs}$, $T_{\rm e} = T_{\rm gas}+(10^4 {\rm K} -T_{\rm gas}) (r/R_{\rm cs}-1)$ if $R_{\rm cs} \leq r \leq 2R_{\rm cs}$ and $T_{\rm e} = 10^4 {\rm K}$ if $r > R_{\rm cs}$.

For the modelling purposes, we vary three values, $Q_{\rm H_2O}$, \tgas{} and ${\rm CH_3OH/H_2O}$ over the logarithmic grids with 10 bins for the each value, and calculate 1000 models for the comet in spring and the same for summer. All the parameters we use in the modelling are given in Table~\ref{tab:model_values}. To aid visualization, we display the parameter space explored for the summer comet simulation in Fig.~\ref{fig:haser_models_summer}.

\begin{table}
    \centering
    \begin{tabular}{c|c}
    \hline
    Parameter & Value \\ 
    \hline
    $Q_{\rm H_2O}$  & $(0.005-5)\times 10^{29}$~s$^{-1}$\\
    \tgas           & $50-1000$~K\\
    ${\rm CH_3OH/H_2O}$ & $0.005-0.5$ \\
    $v_{\rm exp}$ & $0.3$~\kms\\
    $r_{\rm h}$ & 3.4 and 2.7~AU\\
    $\Delta$    & 3.1 and 1.8~AU\\
    \hline
    \end{tabular}
    \caption{Parameters for the model comet. Value of $v_{\rm exp}$ was fixed. The values of $r_{\rm h}$ and $\Delta$ we taken from Table~\ref{tab:detected} for April and July as spring and summer values, respectively. }
    \label{tab:model_values}
\end{table}

Subsequently, we modelled the radiative transfer of the observed methanol line series at 96 GHz and 241~GHz. While water is the most abundant collisional partner in the coma, there are no calculated collision coefficients for \meth{} with water. We follow \citet{2022A&A...660A.118B} using collision coefficients for \meth{} with H$_2$ instead and take the coefficients from \citet{2024MNRAS.527.2209D}. We also consider collisions of \meth{} with electrons from \citet{2013A&A...560A..73G}.

Using the derived spatial distributions of the parameters, we generated synthetic emission maps of the lines belonging to the $2_K-1_K$ and $5_K-4_K$ line series and convolved them with the beam profiles of the OSO~20-m and APEX 12-m telescopes, respectively. While we calculated the molecular line spectra maps for the whole source, we could confront only the simulated and observed spectra toward the centre of the comet since our observations did not spatially resolve the structure of the comet. The beam convolution of our observations corresponds to 
cometocentric radii $\approx 40000 - 90000$~km, i.e. the most of the coma material falls into one beam. 
We note that using the only central convolved spectrum, we do not restrict ourselves to the central part of the coma since the spectrum is formed by the whole coma.

In the paper, we did not adopt the commonly used reduced $\chi^2$-criterion, $\chi^2 = 1/N \times \sum_i ({\rm Obs}_i - {\rm Theory}_i)^2/\sigma_i^2$, where  $\sigma_i$ is the observational noise. The problem of using the reduced $\chi^2$ is that the intensities of our lines differ greatly in magnitude so the reduced $\chi^2$ is totally dominated by the only line with highest intensity. We probed other criteria which would be sensitive to all the lines simultaneously and decided to use the criterion from a book ‘Practical Statistics for Astronomers’ \citep[][]{Wall_Jenkins_2012}. The latter criterion is also
referred as $\chi^2$ but is given by the formula: $\chi^2 = \sum_i ({\rm O}_i - {\rm E}_i)^2/{\rm E}_i$, where ${\rm E}_i$ and ${\rm O}_i$ are the "expected" and "observed" values, respectively. We used our observed integrated intensities as the "observed" values and simulated values as the "expected" ones. This criterion gives the $\chi^2$ value to be close to zero for a good fit and is sensitive to all the lines at the same time.

\subsection{Results of non-LTE modelling}

\begin{figure*}
    \centering
    \includegraphics[width=0.45\linewidth]{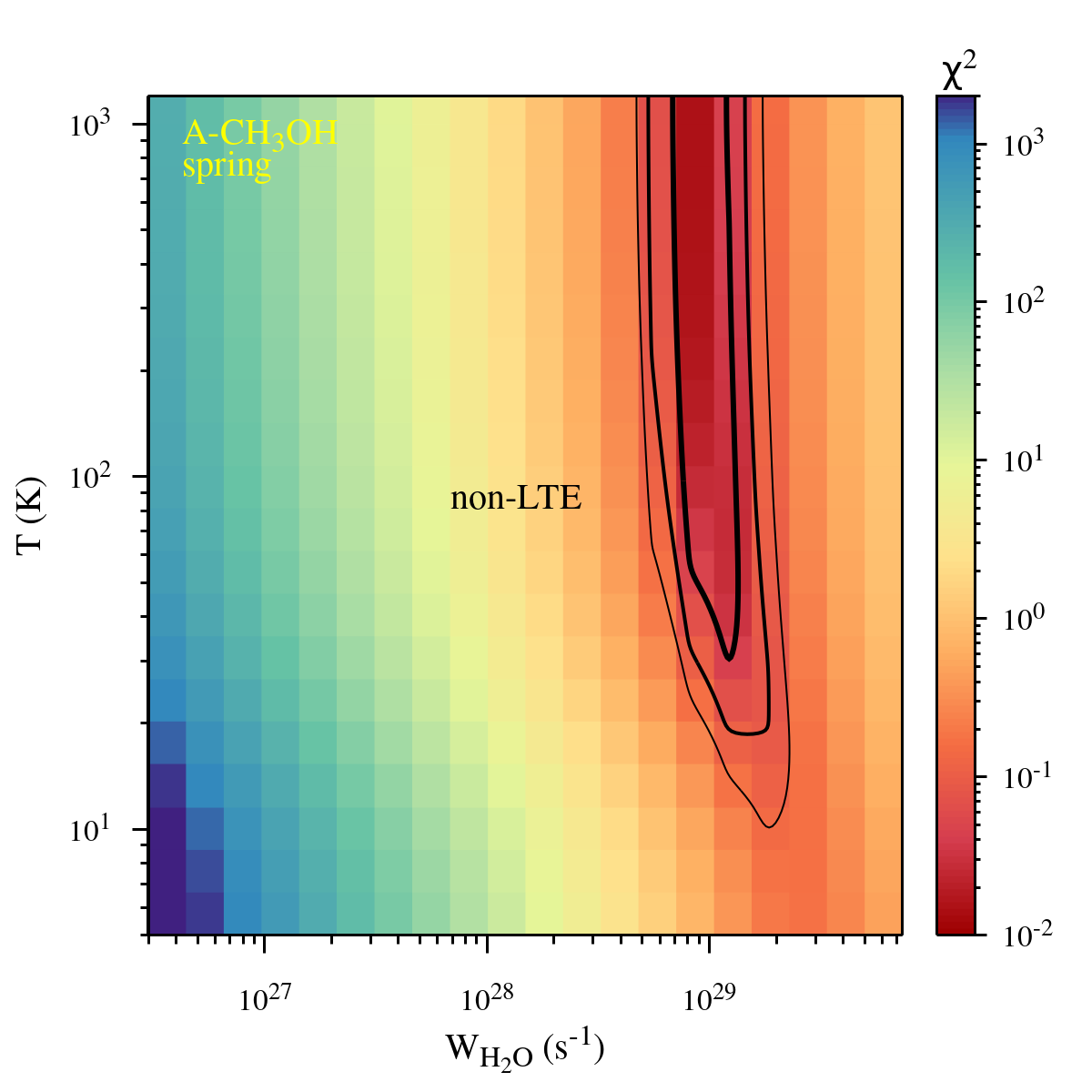}
    \includegraphics[width=0.45\linewidth]{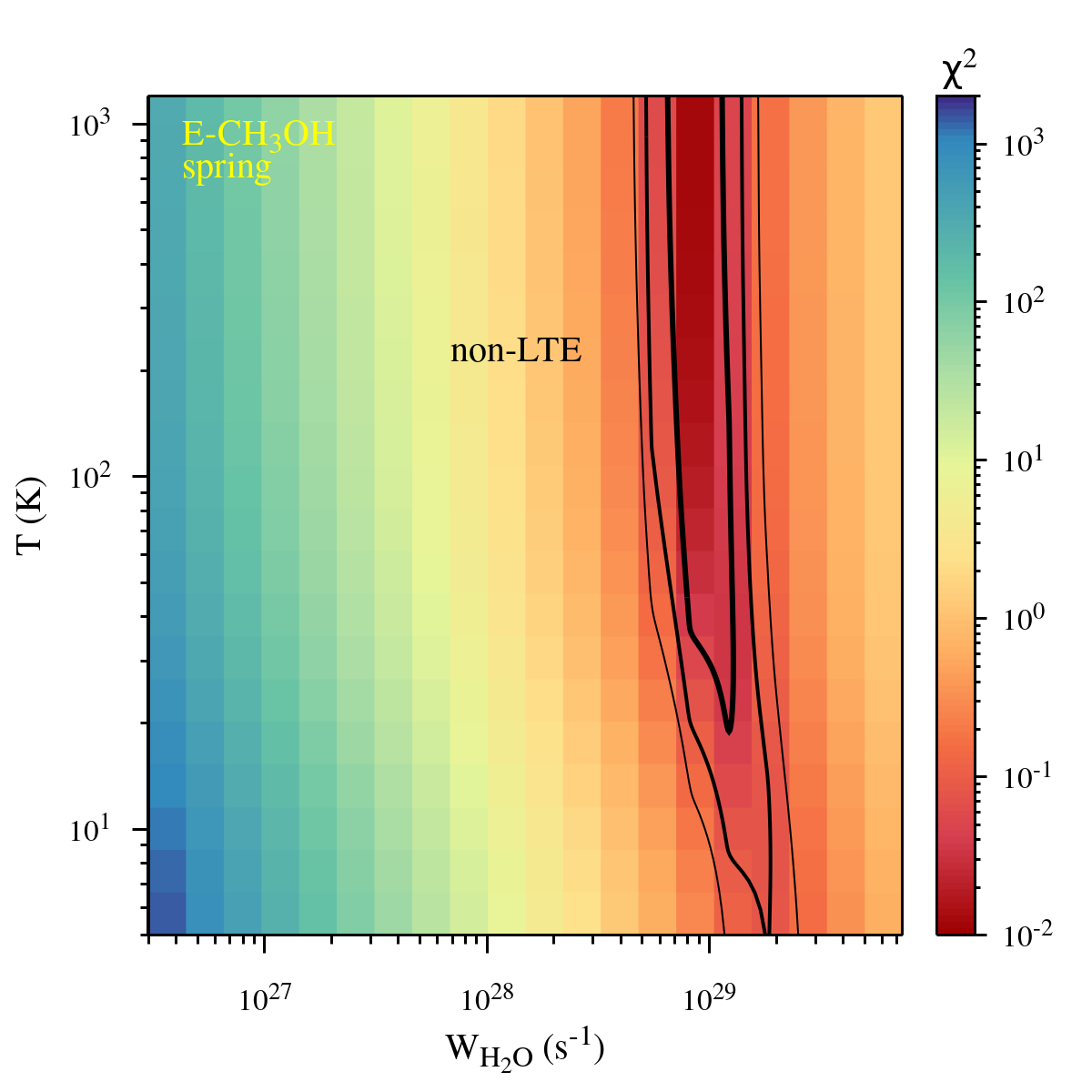}\\
    \includegraphics[width=0.45\linewidth]{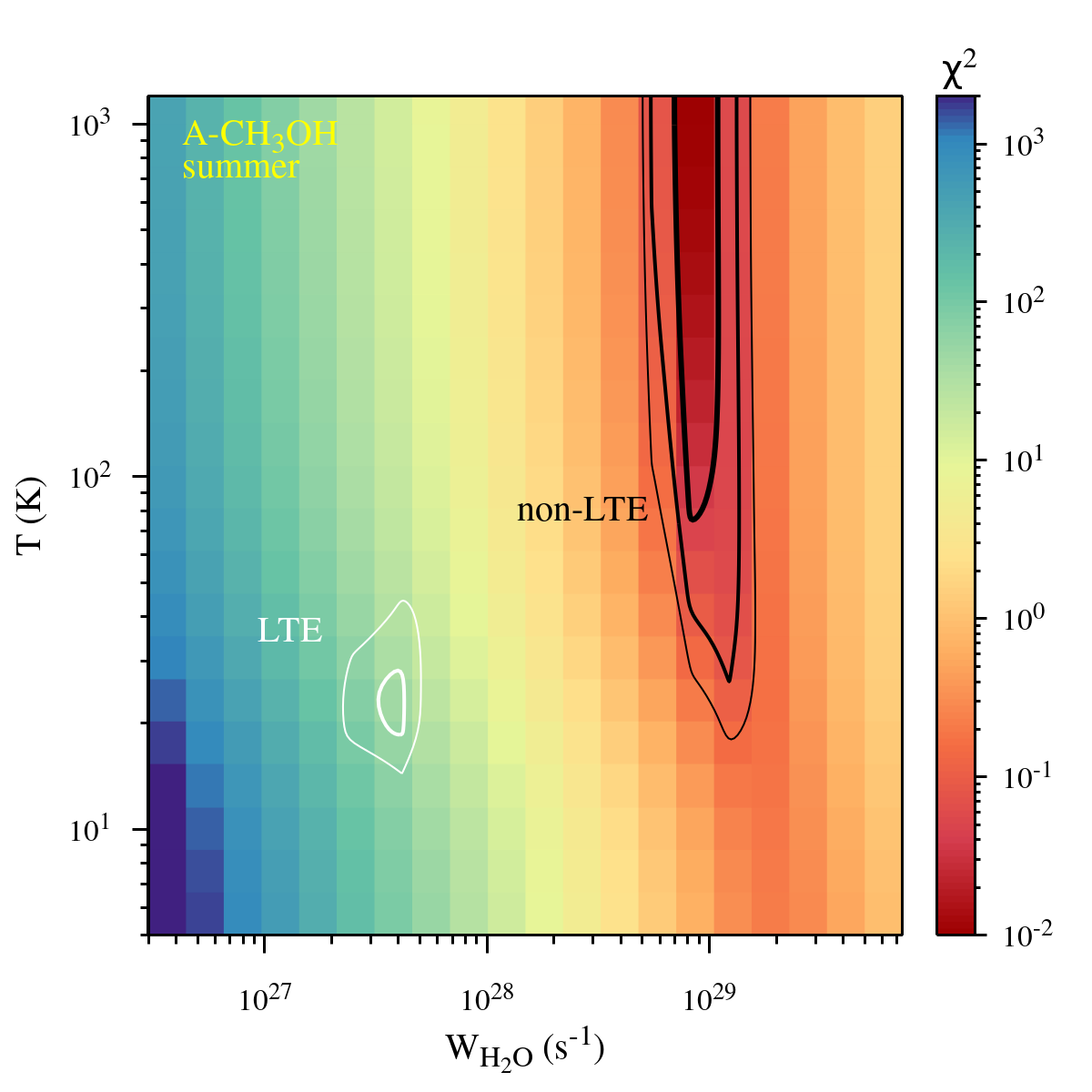}
    \includegraphics[width=0.45\linewidth]{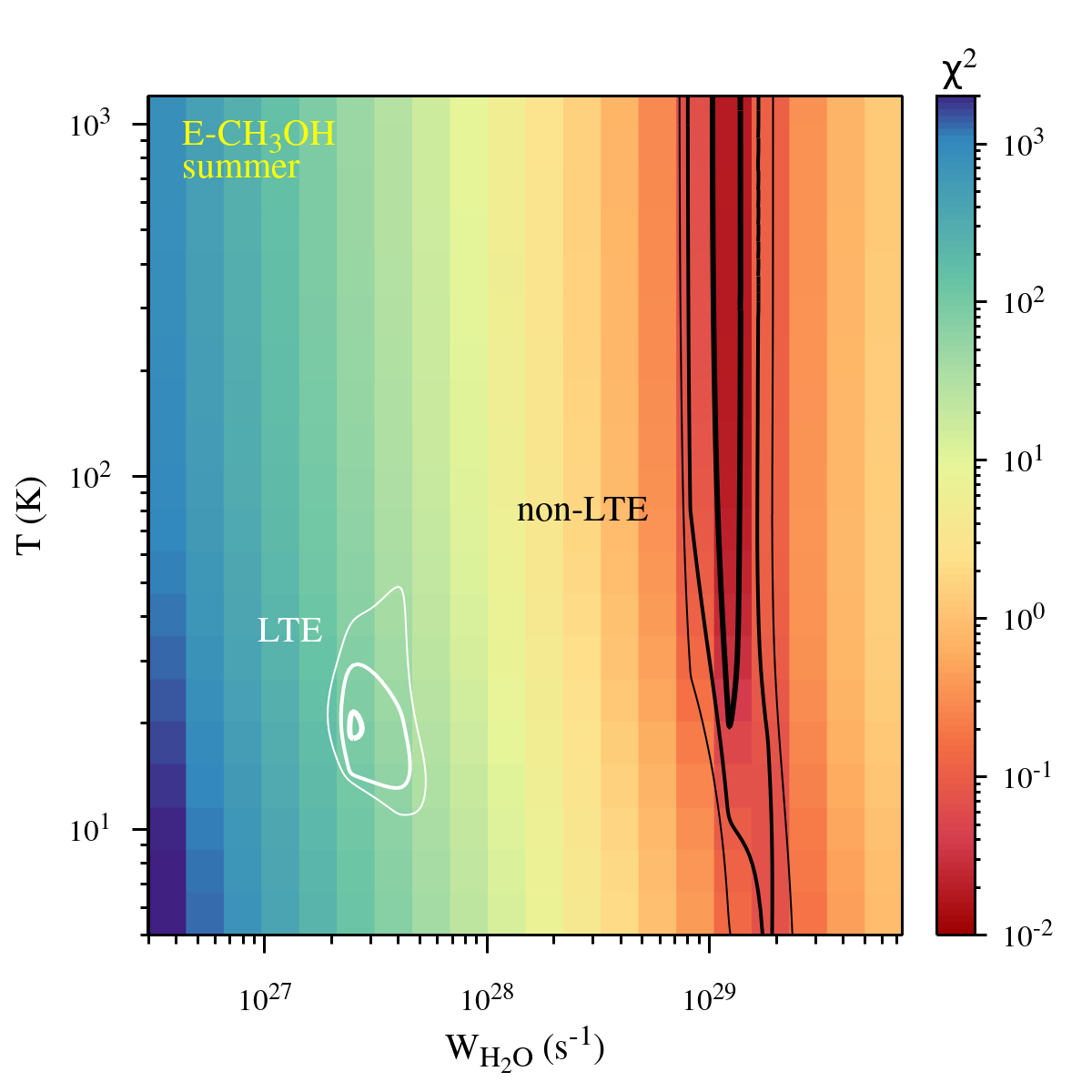}
    \caption{Maps of the $\chi^2$ criteria (colour) for the spring (top row) and summer (bottom row) observations. The intervals are shown by black lines. The maps are shown for those values of the $Q_{\rm CH_3OH}/Q_{\rm H_2O}$ ratio providing the minimum of the $\chi^2$ values (see text). The white lines on the bottom panel shows the parameter space where the LTE calculations give the best fit. The contours correspond to the 0.04, 0.08 and 0.13 values of the $\chi^2$.}
    \label{fig:best_fit_models_cubes}
\end{figure*}

The non-LTE modeling of the methanol emission was performed separately for the A and E-modifications as they have different collisional cross-sections. Parameters of the best-fit model ($Q_{\rm H_2O}$, ${\rm CH_3OH/H_2O}$ and \tgas) are given in Table~\ref{tab:chi_square_results}. We note the broad range, spanning more than for two orders of magnitude, of the gas temperatures in coma. We also pay attention to the high values of \tgas{} for the best-fit models, namely $T_{\rm gas} \geq 250$~K. These values are $40-70$ times higher that we obtained doing the LTE analysis for spring and summer. Two-dimension distribution of the $\chi^2$ criteria over the ${\rm CH_3OH/H_2O}$ and \tgas{} ranges for the fixed $Q_{\rm H_2O}$ from the best-fit model can be seen in Fig.~\ref{fig:best_fit_models_cubes}. We conclude that the best-fit values of the parameters are about $Q_{\rm H_2O} = (3-10)\times 10^{28}$~s$^{-1}$ for the ${\rm CH_3OH/H_2O} \approx 0.02-0.04$. As the comet was closer to the Sun in summer~2022 than in spring, we reasonably obtained higher values of the water production, whose best-fit values $Q_{\rm H_2O} = (7-10)\times 10^{28}$~s$^{-1}$.  Conversely, about a factor of 2 lower values of the ${\rm CH_3OH/H_2O}$ were found in summer. 

\begin{figure*}
    \centering
    \includegraphics[width=0.45\linewidth]{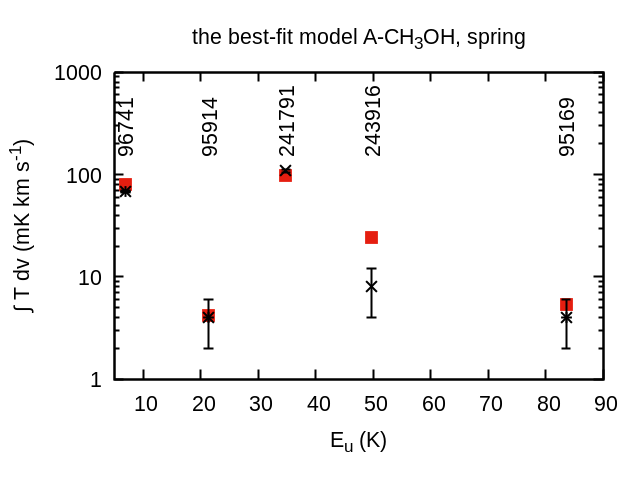}
    \includegraphics[width=0.45\linewidth]{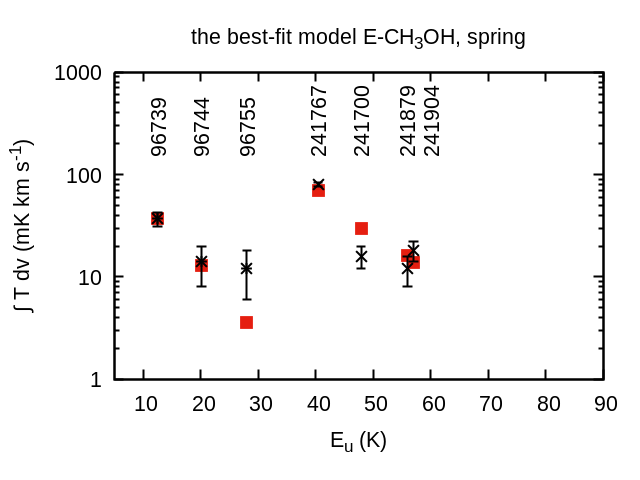}\\
    \includegraphics[width=0.45\linewidth]{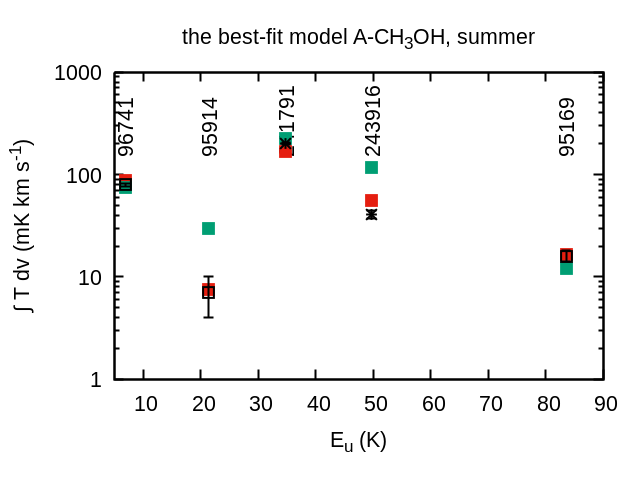}
    \includegraphics[width=0.45\linewidth]{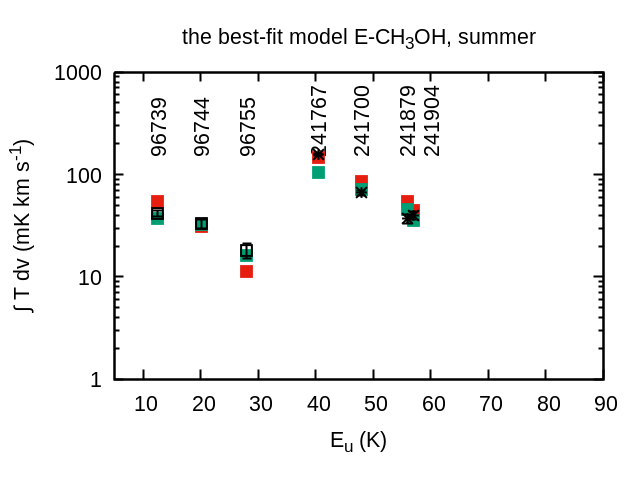}
    \caption{The best-fit models for the spring (top row) and summer (bottom row) observations. Red (green) squares show the non-LTE (LTE) simulations in the bottom row.}
    \label{fig:best_fit_models}
\end{figure*}

Comparison of the observed and simulated integral intensities for the methanol lines is shown in Fig.~\ref{fig:best_fit_models}. In spring, four of five lines of A-methanol and four of seven lines of E-methanol show agreement within the observational uncertainties. The A-methanol line at 243915.788~MHz and E-methanol lines at 241700.159~MHz and 96755.501~MHz show agreement within factors of two-three. In summer, five of five lines of A-methanol and six of seven lines of E-methanol show agreement between the observed and simulated values within the observational uncertainties. Disagreement between the model and observations is highest for the E-line at 96755.501~MHz, but it is less than a factor of two.

\begin{table}
    \centering
    \begin{tabular}{c|c|c|c|c}
    \hline
                            & Sym & $Q_{\rm H_2O}$        & ${\rm CH_3OH/H_2O}$ & \tgas \\
                            &    & ($10^{28}$~s$^{-1}$) & --                  & (K)   \\
    \hline
    \multicolumn{5}{c}{non-LTE}      \\
    \multirow{2}{*}{spring} & A & 3.3                   & 0.044                & 30-1000 (1000)\\
                            & E & 7.1                   & 0.017                & 20-1000 (760) \\
    \multirow{2}{*}{summer} & A & 7.1                   & 0.017                & 80-1000 (1000)\\
                            & E & 10.5                  & 0.013                & 20-1000 (250) \\
    \multicolumn{5}{c}{LTE}\\
    \multirow{2}{*}{summer} & A & 0.4                   & 0.50                 & 23  \\
                            & E & 0.2                   & 0.50                 & 18  \\
    \hline
    \end{tabular}
    \caption{The obtained values of the best-fit model parameters from the non-LTE and LTE calculations. Intervals $\pm \sigma$ for temperature are shown for the non-LTE with the best-fit value shown in brackets.
    }
    \label{tab:chi_square_results}
\end{table}

To investigate the significant discrepancy between the gas temperatures derived from population diagrams and those obtained through the non-LTE approach, we performed LTE calculations using the URAN(IA) code. In these calculations, the excitation temperatures of the methanol emission lines were set equal to the gas temperature. Distribution of $\chi^2$  values is presented in the lower section of the Fig.~\ref{fig:best_fit_models}. Although the parameter space of the best-fit LTE models spans a substantial range of \tgas{} and $Q_{\rm H_2O}$ values, this region shows no overlap with the best-fit parameter space in the non-LTE calculations. This lack of intersection suggests fundamental differences in the underlying physical assumptions between the two approaches. We obtain $T_{\rm gas} \leq 40$~K with the best-fit values $\approx 20$~K for the LTE approach for summer, $Q_{\rm H_2O}$ up to $20-50$ times smaller compared with the non-LTE and ${\rm CH_3OH}/{\rm H_2O} = 0.5$. In spite of the fact that the minimum of the $\chi^2$ values were found for substantially different model parameters, we see reasonable agreement between the simulated and observed integrated intensities in Fig.~\ref{fig:best_fit_models} for the summer. Only for two lines of A-methanol we see difference of the half order of magnitude. While both non-LTE and LTE intensities agree with the observed line intensities in the non-LTE and LTE models, the best-fit physical parameters in these two approaches are different. Since the non-LTE approach is more general than the LTE, we consider the non-LTE parameters to be closer to the physical conditions in the comae. 

\begin{figure*}
    \centering
    \includegraphics[width=0.45\linewidth]{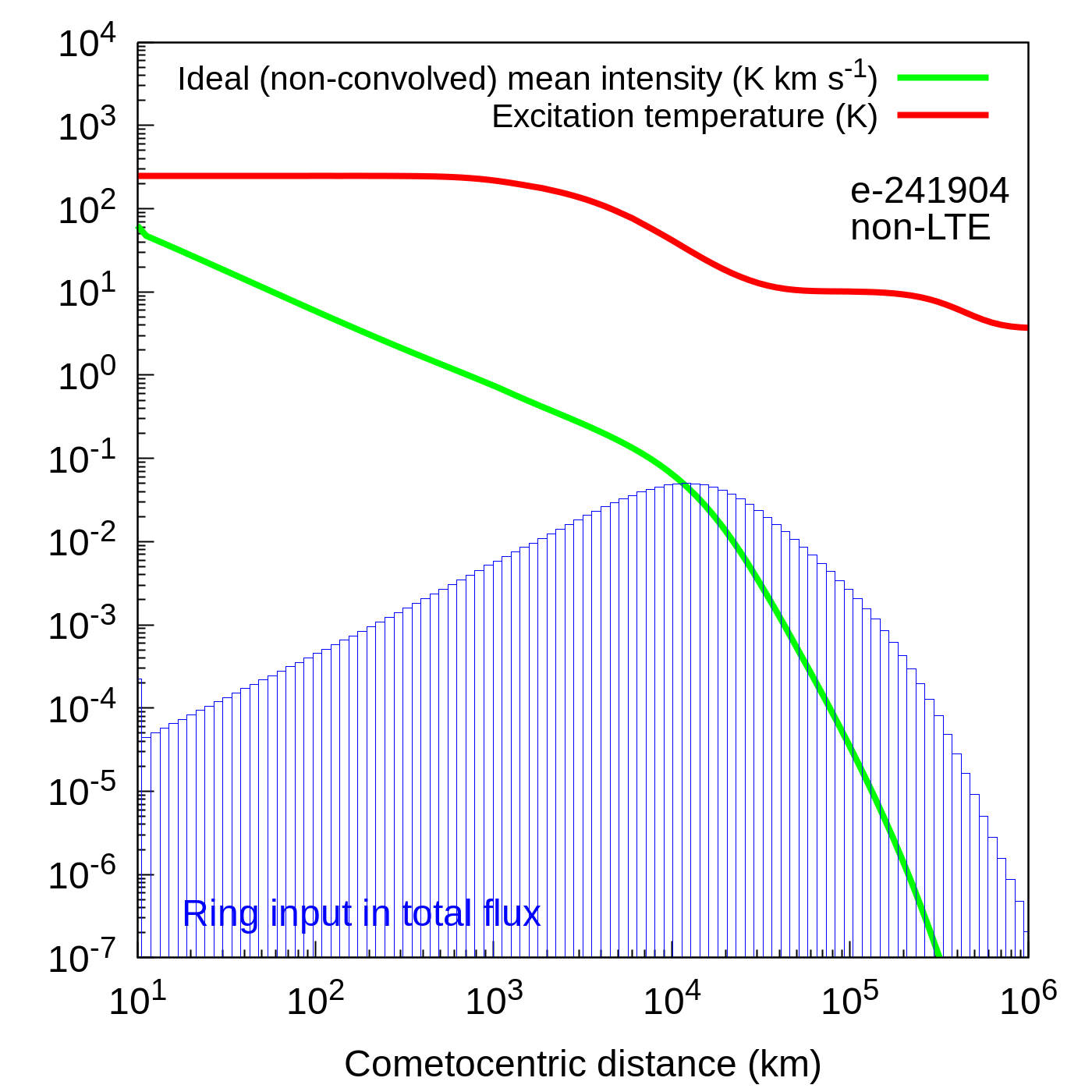}
    \includegraphics[width=0.45\linewidth]{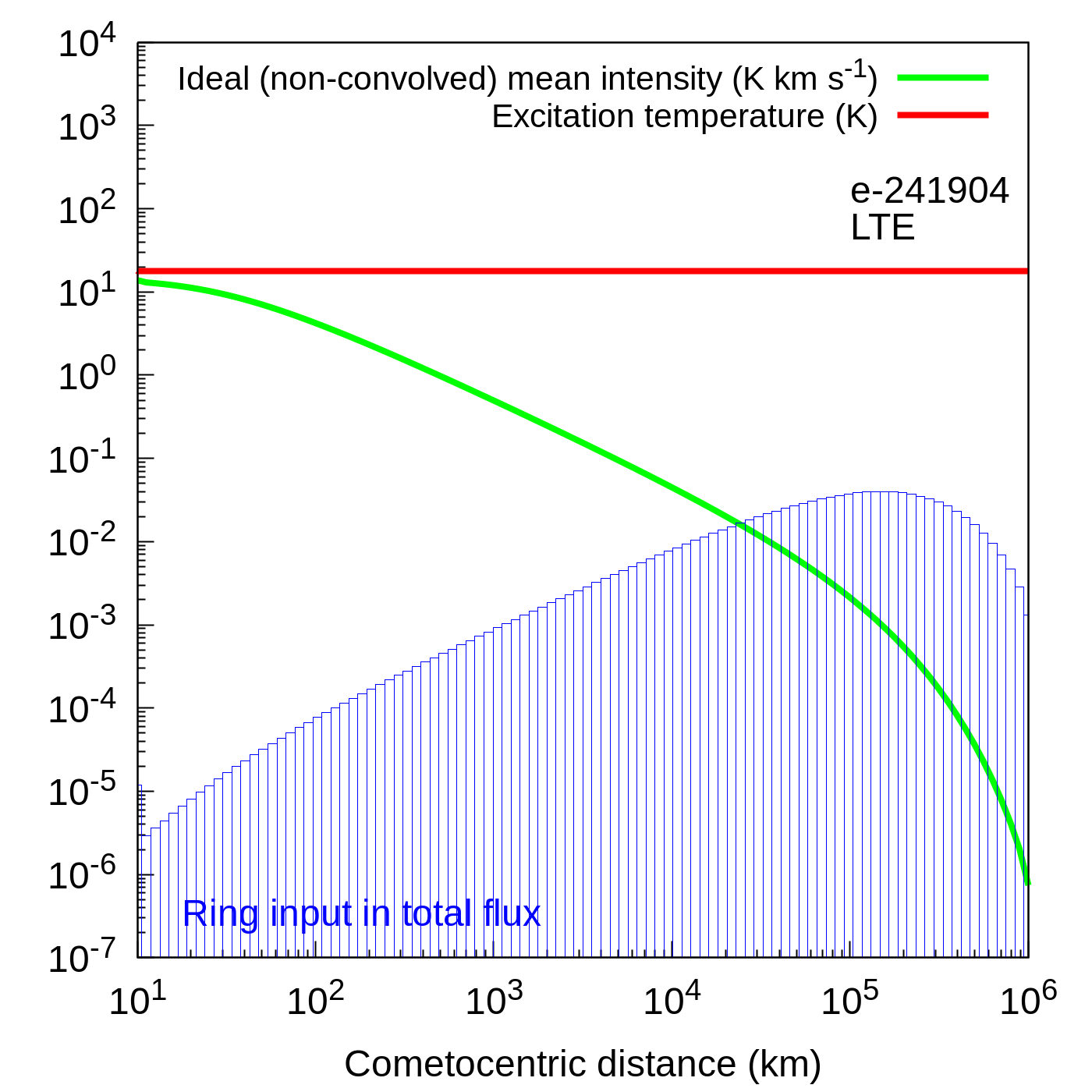}\\
    \caption{The results of radiative transfer simulations for the best-fit non-LTE (left panel) and LTE (right panel) models.
    The spatial distribution of excitation temperature of the E-methanol line at {241904}~MHz is shown by the red colour. The emergent non-convolved mean intensity distributions as a function of impact parameter are shown by green. The blue boxes represent the contribution of every pre-selected ring (within the box boundaries) in the image plane into the total line flux from the source.}
    \label{fig:dilution}
\end{figure*}

In order to understand why we obtain the reasonable fits of the observed lines with both the non-LTE and LTE models we provide Fig.~\ref{fig:dilution} with the radial distributions of the simulated excitation temperature, calculated for the best-fit non-LTE (left panel) and LTE (right panel) summer models. The plot also includes the spatial distributions of the emergent non-convolved mean intensity (green lines) and the contribution (blue bars) of the pre-selected rings in the image plane (with different impact radii around the source centre) to the total flux. 

We see that spatial distributions of the emergent non-convolved mean intensity are quite different in LTE and non-LTE cases. In particular, the LTE mean intensity distribution is more spatially extended. Given the spatially resolved observations, it would be be possible to distinguish between the models. However, the beam of our observations is comparable with the considered extent of the comet (note the spatial resolution of our observations in Table~\ref{tab:detected}). Therefore, we have to compare the emergent total line fluxes which appear to be similar in both models. 

The excitation temperature in the LTE model follows the kinetic gas temperature and is equal to 20~K over the whole source. In the LTE-model, the contribution of the particular concentric domain to the total flux is highest at the cometocentric distance of $10^5$~km. 
In the non-LTE model, the excitation temperature for the selected transition varies from $\approx$200~K near the centre to $\approx$5~K at the outer edge but the highest contribution into the total flux provides the region with cometocentric distance of $10^4$~km where the excitation temperature is $\approx$20~K. So, in both models the total flux is mainly contributed by the emission from the regions with the same excitation temperatures.

In spite that both non-LTE and LTE models provide reasonably good fit to the data,  our best-fit non-LTE values of $Q_{\rm H_2O}$ and ${\rm CH_3OH}/{\rm H_2O}$ are remarkably different compared with the LTE results. 
Therefore, we conclude that LTE analysis of the integral source emission (i.e. total flux) may give incorrect physical parameters of the object.  

\begin{figure*}
    \centering
    \includegraphics[width=0.45\linewidth]{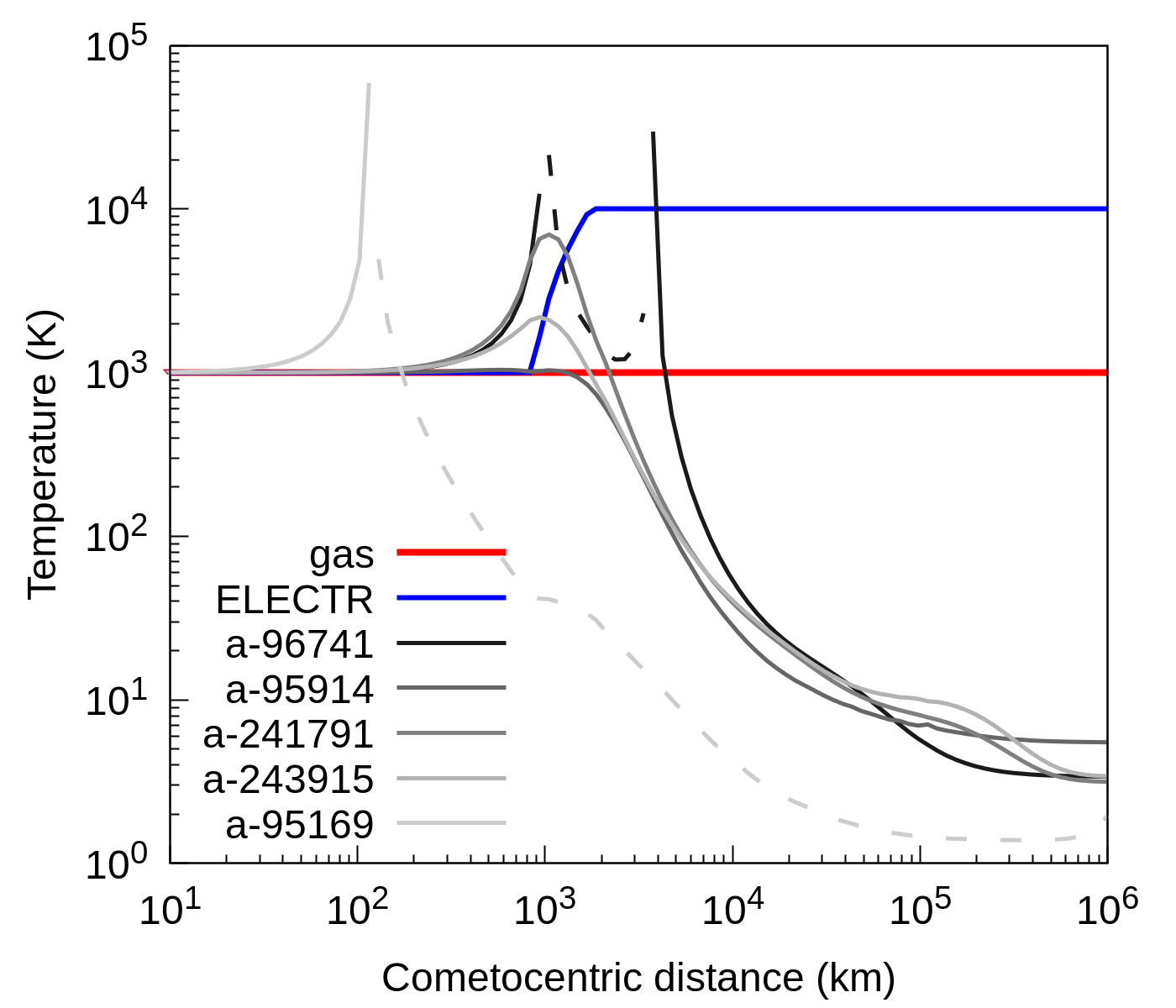}
    \includegraphics[width=0.45\linewidth]{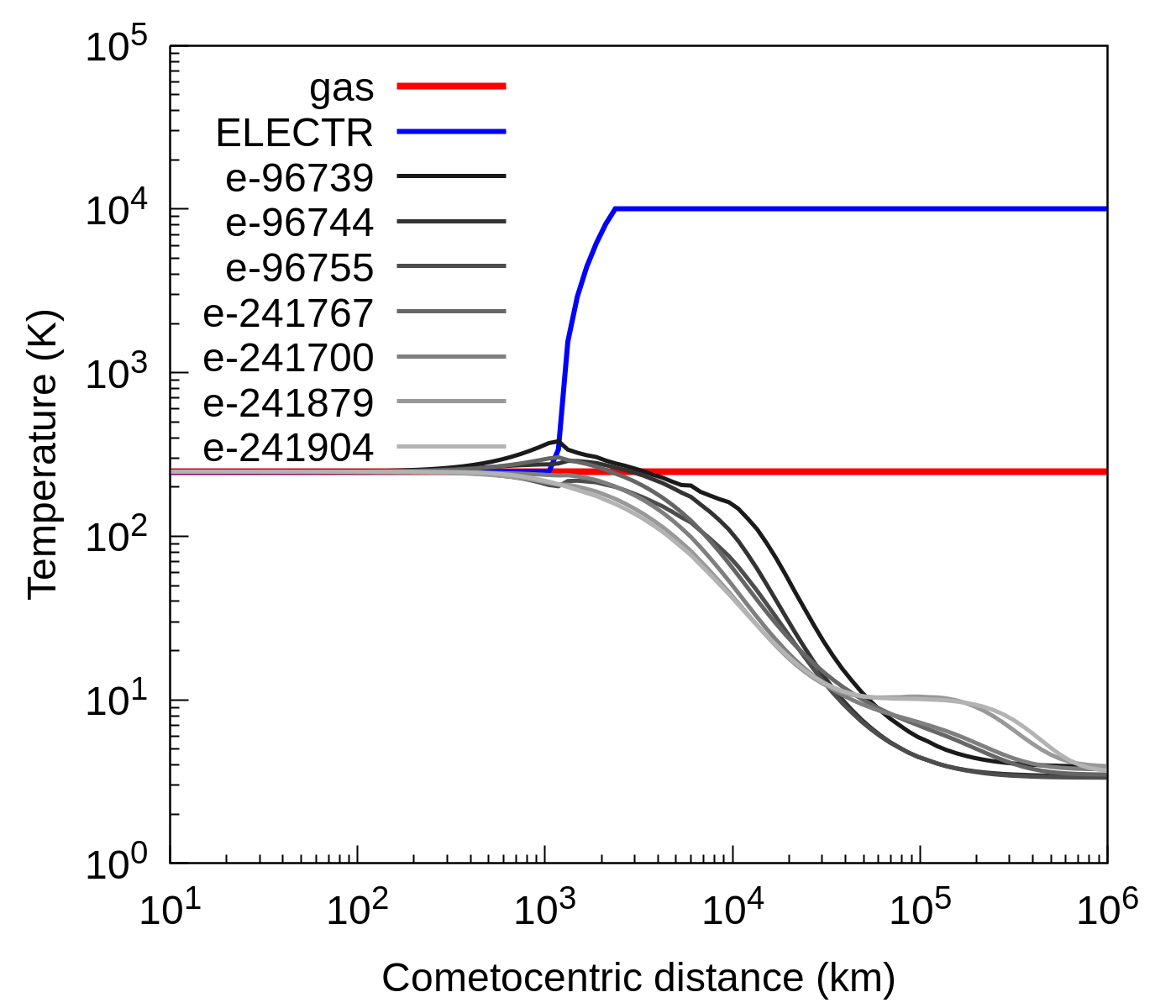}
    \caption{Temperature distribution in the coma of the comet. The gas temperature, electron temperature and excitation temperatures of the observed methanol lines are shown by red, blue and grey lines, respectively. Dashed line corresponds to negative values of the excitation temperatures. the frequencies of the lines are given in MHz.}
    \label{fig:t_ex}
\end{figure*}

Excitation conditions for the methanol lines of the $2_K-1_K$ and $5_K-4_K$ series are far from equilibrium in our calculations, as can be seen in Fig.~\ref{fig:t_ex}, where we show excitation temperature of the lines in the best-fit models for the summer. While for E-methanol the difference between the \tgas{} and the excitation temperature values do not exceed factor of two for the cometocentric distance of $10^3-10^4$~km, it reaches two orders of magnitude in the outer part of the comae. For A-methanol, deviation of the excitation conditions from the equilibrium is ever higher. Namely, we find negative values of the excitation temperature for the $8_0-7_1$ and $2_0-1_0$ lines at 95169.391 and 241791.352~MHz, respectively. The cometocentric velocity of the $8_0-7_1$ line is different from all the methanol lines observed in July. Probably, the maser conditions do not appear in the whole coma but locally. Therefore, we may find weak maser emission of methanol in the comet.

\section{Discussion}\label{sec:disc}

Non-LTE analysis of the molecular line emission has become a standard approach for interpreting single-dish observations of cometary comae \citep[see e.~g.][]{LEE2011721, 2022A&A...660A.118B}. 

Checking the literature, we found the maser emission of water lines in comets at 22 GHz was theoretically predicted by \citet{1983SvAL....9...99S} based on observations by \citet{1981A&A....97..195C}. Although several subsequent attempts were made to detect molecular maser emission in comets \citep[e.g.,][]{2000AJ....119.2465G, 2014P&SS...96...22C, 2020EPSC...14..171S, 2024RAA....24j5008C}, this area of research has not been extensively developed. To our knowledge, methanol masers have not been detected in comets. \citet{1992MNRAS.259..203C} calculated the excitation conditions for methanol and predicted that the $8_0-7_1$ line could exhibit maser activity under typical physical conditions in star-forming regions. The pumping mechanism for this line involves collisional excitation followed by spontaneous radiative decay, classifying it as a Class I methanol maser transition. We suggest that this research area could be further explored in the future, given that mm-wave emission at frequencies around 90 GHz is routinely observable from Earth. The methanol excitation and the maser phenomena are better explored using improved molecular data, e.g., collisional rate coefficients extended for the high energy transitions as it was shown \citep[e.~g.][]{1983PAZh....9...26S, 2005MNRAS.360..533C, Kalenskii2016, 2025INASR..10...10F}.

Comparing our values of $Q_{\rm H_2O}$ with results obtained by \citet{2023A&A...674A.206K, 2025AJ....169..102E, 2025PSJ.....6..139W},  we find an agreement between their non-LTE estimations and ours. Indeed, \citet{2025AJ....169..102E} report about $1.2\times10^{28}$\,s$^{-1}$, $2.4\times10^{28}$\,s$^{-1}$ and $3.6\times10^{28}$\,s$^{-1}$ for the heliocentric distances 2.54, 2.48 and 2.35~AU, respectively. \citet{2023A&A...674A.206K} report the upper limit for $Q_{\rm H_2O}$ of $7\times10^{28}$\,s$^{-1}$. \citet{2025PSJ.....6..139W} performed spatially resolved observations of the comet using JWST and obtained model-derived values $(2-4)\times10^{28}$\,s$^{-1}$ for the 6~um water band and up to $8\times10^{28}$\,s$^{-1}$ for the hot 5\,um band. It was shown that $Q_{\rm H_2O}$ becomes higher at higher cometocentric distances. They also mention the much larger $Q_{\rm H_2O}$ up to $2\times10^{29}$\,s$^{-1}$ deduced from the OH emission line at 18\,cm obtained with the larger beam size of the Nancay telescope. They suggest an extended source of water in the coma for this particular comet. The low limit of our estimations of $Q_{\rm H_2O}$ agrees with the results of \citet{2023A&A...674A.206K} and \citet{2025PSJ.....6..139W}. Moreover, our best-fit model for summer agrees with \citet{2025PSJ.....6..139W}.

However, their values of ${\rm CH_3OH/H_2O}$ are higher than our best-fit non-LTE values by 5-6 times. The ${\rm CH_3OH/H_2O}$ ratio from the spatially-resolved analysis by \citet{2025PSJ.....6..139W} is about 2-3 times higher than our best-fit values. Refraining from far-reaching conclusions about these differences, we note that selected 12 methanol lines became not the best set to definitively determine the $Q_{\rm H_2O}$, ${\rm CH_3OH/H_2O}$ and $T_{\rm gas}$ quantities. The combination of other series of the methanol lines probably improves the situation, as suggested by \citet{Kalenskii2016}. We note that the temperature uncertainty we found can be related to the extended region of methanol production and emission in the coma, as was explored by \citet{2025arXiv251105662R} using spatially-resolved ALMA data. 

The comet crossed the water ice sublimation line during our observation sessions in spring and summer of 2022. The methanol line emission became about $1.5-2$ times brighter in summer due to the approaching of the comet to the Sun. However, the brightness of the CO and HCN lines was not changed. We suggest this result can be related with the depletion of these molecules from the comet's ices. First, the activity of C/2017~K2 was driven by the sublimation of a super-volatile CO ice at $r_{\rm h} > 10-20$~AU as was reported by \citet{2017ApJ...849L...8M} and \citet{2019AJ....157...65J}. Astrochemical models of the protoplanetary disks predict that outer disk regions, including the protosolar one, are enriched by CO but the water-rich region is closer to the star \citep[see e.~g.][]{2023A&A...670A..28S, 2024MNRAS.530.2731T}. Second, \citet{2021PhR...893....1O} summarises results of astrochemical modelling of the last decades and discusses a model of layered ice mantle of the interstellar grain, where the bulk made of water is covered by a surface enriched by CO, CO$_2$ and CH$_3$OH. Evidence of the evaporation of the mantle top layer at the beginning of the mantle destruction under the UV radiation can be found in young star-forming regions \citep[see e.~g.][]{2025ABULL}. Even outbursts of the stellar luminosity, peculiar to young stars, can not erase this structure completely, see \cite{2022ARep...66..393B}. Therefore, the CO depletion can be attributed to the disappearing of the top layer of the grain mantle. Third, the depletion of a super-volatile CO ice can be related to previous perihelion passage of the comet reported by \citet{2018A&A...615A.170K, 2018RNAAS...2...10D}, while \citet{2022DPS....5441103B} argue that C/2017~K2 is on its first close approach to the Sun. 

Another way to explain the different behaviour of the CO, HCN and methanol line intensities is to propose the existence of more then one source of outgassing molecules. Different line widths of the CO, HCN and methanol lines found by us suggest that the outgassing velocity might be different due to anisotropic spatial distribution of the molecules around the nucleus. For example, \citet{Gunnarsson03} found an insolation-dependent and also isotropically outgassing sources of molecules observing comet C/1995~O1 with single-dish SEST telescope. \citet{Biver19} found a narrow jet outgassing and more isotropic source analysing the MIRO submillimeter radiotelescope data of comet 67P. \citet{2021A&A...651A..29I} studied C/2011~KP36 and found several large active areas consisting of different combinations of water ice, CO$_2$ ice, and refractory dust. We probably deals with similar phenomenon in~C/2017~K2.

\section{Conclusions}

We observed 12 lines of methanol, one line of CO and one of HCN in the Oort comet C/2017~K2~(PANSTARRS) during the pre-perihelion phase from April to July 2022 with the APEX 12-m and OSO 20-m telescopes. The study demonstrates the necessity of multi-line, non-LTE analyses to accurately derive physical and chemical conditions in cometary atmospheres. Our conclusions are following.

\begin{itemize}
    \item Methanol line intensities increased significantly (by 10–100\% or more) from spring to summer, particularly for transitions with upper energy levels $E_{\rm u} > 40$~K. In contrast, CO and HCN line brightness did not increase over the same period. The coma expansion velocity remained constant at 0.3~km~s$^{-1}$.

     \item A preliminary LTE analysis using rotation diagrams suggested rotational temperatures of $T_{\rm rot} = 10-20$~K. However, non-LTE radiative transfer modeling revealed significantly higher gas temperatures, namely $T_{\rm gas} > 100$~K (withthe best-fit values $T_{\rm gas} = 250-1000$~K) and water production rates $Q_{\rm H_2O}=(3-10)\times10^{28}$~s$^{-1}$. The methanol-to-water abundance ratio was found to be $\approx 0.02-0.04$ in spring and $\approx 0.01-0.02$ in summer. The $8_0-7_0$ methanol line showed signs of non-LTE excitation, including possible weak maser activity.  The discrepancy between LTE and non-LTE derived parameters underscores the importance of non-LTE modeling for cometary comae. The potential detection of methanol maser emission highlights the complex excitation conditions in comets and opens a new avenue for future research.

     \item The principal outcome of our paper is the warning that LTE-analysis of the studied comet molecular emission may be misleading since the results of LTE-simulatons are not reproduced by more general (and more correct) non-LTE simulations. On the other hand, the non-LTE simulations did not allow us to isolate the best-fit model due to the strong parameter degeneracy.

    \item The results imply that C/2017 K2 may have already lost its most volatile ices before perihelion, supporting the hypothesis that it is not on its first approach to the Sun. The increase in methanol emission with decreasing heliocentric distance is consistent with enhanced sublimation as the comet approached the Sun. The lack of increase in CO and HCN emission suggests depletion of these volatiles, likely due to previous sublimation at larger distances or the loss of a surface layer enriched in CO and CO$_2$. Alternatively, the different behavior of the CO, HCN, and methanol line intensities could be related to the existence of more than one source of outgassing molecules.

\end{itemize}

\section*{Acknowledgements}

We are thankful to O.~Ivanova, who inspired us for the study and advised to observe the particular comet. We also thank S.~V.~Kalenskii, S.~V.~Salii, B.~M.~Shustov and A.~M.~Sobolev for helpful discussions and anonymous referee for his/her questions and suggestions.

This research has made use of spectroscopic and collisional data from the EMAA database (https://emaa.osug.fr and https://dx.doi.org/10.17178/EMAA). EMAA is supported by the Observatoire des Sciences de l’Univers de Grenoble (OSUG). 

\section*{Data Availability}

The data are available upon reasonable request of the authors.



\bibliographystyle{mnras}
\bibliography{K2_PANSTARRS} 




\appendix

\section{Ancillary figures}

\begin{figure}
    \centering
    \includegraphics[width=\columnwidth]{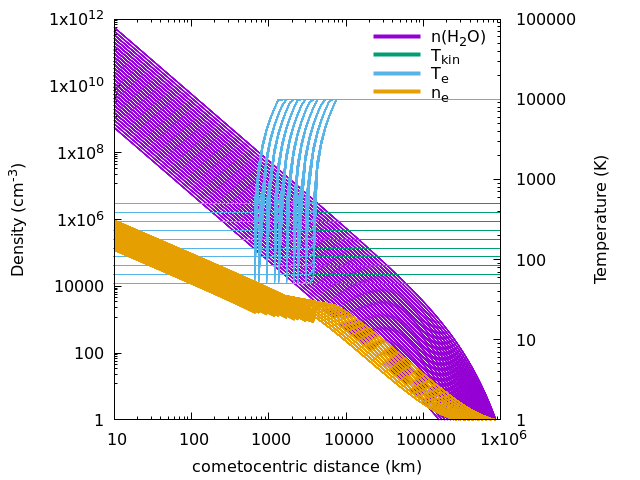}
    \caption{Parameter space explored for the summer comet simulation: neutral gas temperature (green), electron temperature (blue), water and electron number densities (magenta and orange, respectively).}
    \label{fig:haser_models_summer}
\end{figure}

\begin{figure}
    \centering
    \includegraphics[width=0.9\linewidth]{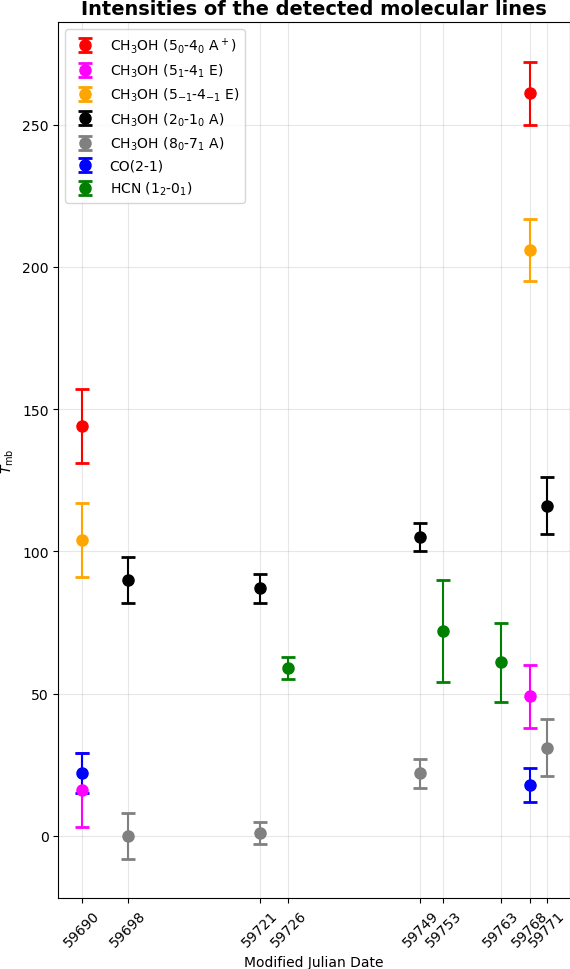}
    \caption{Intensities of several detected molecular lines over the particular period of Modified Julian Date (MJD) we observed the comet.}
    \label{fig:julian_date}
\end{figure}

\bsp	
\label{lastpage}
\end{document}